\documentclass[aps,
  prl,
  reprint,
  longbibliography,
  amsmath,
  amssymb,
  tightenlines，
]{revtex4-2}

\usepackage{graphicx}
\usepackage{epstopdf}
\usepackage{dcolumn}
\usepackage{bm}
\usepackage{lipsum}
\usepackage{booktabs}

\usepackage{hyperref}
\usepackage[capitalize]{cleveref}

\begin{document}
\preprint{APS/123-QED}

\title{Decoherence Tuning in Voltage-Driven Plasmonic Cavities}

\author{Yuchen Wang\textsuperscript{$\dagger$}}
\altaffiliation{Current affiliation: State Key Laboratory of Precision and Intelligent Chemistry, University of Science and Technology of China, Hefei, China}

\affiliation{Department of Chemistry, New York University, New York, NY 10003}
\affiliation{The Simons Center for Computational Physical Chemistry, New York University, New York, NY 10003}

\author{Oliver Tan}
\altaffiliation{Y.W. and O.T. contributed equally to this work.}
\affiliation{Department of Chemistry, New York University, New York, NY 10003}

\author{Norah M. Hoffmann}
\altaffiliation{Corresponding author: \href{mailto:norah.hoffmann@nyu.edu}{norah.hoffmann@nyu.edu}}
\affiliation{Department of Chemistry, New York University, New York, NY 10003}
\affiliation{The Simons Center for Computational Physical Chemistry, New York University, New York, NY 10003}
\affiliation{Department of Physics, New York University, New York, NY 10003}

\begin{abstract}

\noindent Strong light-matter coupling offers a powerful framework for modifying molecular structure, dynamics, and optical response, yet achieving this regime at the single-molecule level remains a central challenge. Scanning Tunneling Microscope Break Junctions (STM-BJ) provide a promising route by combining single-molecule addressability with extreme nanoscale field confinement. However, the metallic environment that enables such confinement also introduces substantial plasmonic and excitonic losses, raising the question of how strong coupling can be realized in a metal-molecule-metal junction. Here, we identify the voltage as the key control parameter. We show that the applied bias not only drives the formation of an interfacial exciton through resonant charge transport but also suppresses its coupling to metallic loss channels. Simultaneously, the plasmonic cavity retains extreme mode confinement while the availability of modes is moderately tuned by voltage. Together, these results establish voltage-driven STM-BJs as electrically tunable single-molecule plasmonic cavities and open new routes for using the powerful STM-BJ toolbox, including quantum transport, tracking chemical dynamics, optoelectronics, and molecular-scale quantum control under strong light-matter coupling.

\end{abstract}

\maketitle

\section*{Introduction}

Strong light–matter coupling arises when confined electromagnetic modes hybridize with material excitations faster than the relevant dissipative rates, yielding polaritonic states with a mixed light–matter character.\cite{dovzhenko2018light, ballarini2019polaritonics, rivera2020light,wei2021plasmon} This regime has enabled new routes across applications ranging from chemical reactivity, \cite{thomas2016ground,thomas2019tilting,ahn2023modification, hutchison2012modifying,herrera2016cavity} enhanced exciton transport, \cite{rozenman2018long,pandya2022tuning} long-range energy transfer \cite{coles2014polariton,du2018theory,delpo2021polariton} to room-temperature polariton condensation,\cite{kasprzak2006bose,kena2010room} and tuning of phase transitions.\cite{wang2014phase}

These strong light-matter coupling platforms commonly rely on collective coupling, where many emitters couple simultaneously to a cavity mode. Although this collective enhancement is central to many experiments, ensemble averaging, molecular disorder, and dark-state manifolds can make it difficult to achieve microscopic interpretability, deterministic control, and molecular-scale functionality.\cite{ying2025collective, herrera2020molecular, sandik2025cavity,Gera2022Disorder} Hence, an alternative route is to increase the coupling strength by reducing the electromagnetic mode volume rather than increasing the number of molecules. Plasmonic cavities are therefore attractive because they can confine optical modes below the diffraction limit\cite{lee2023strong} and to nanoscale volumes,\cite{bitton2022plasmonic,torma2014strong} enabling strong coupling with few or single molecules. \cite{chikkaraddy2016single,paoletta2024plasmon}

However, this strong confinement comes at the cost of increased loss, as confined optical modes are located near metallic environments.\cite{khurgin2012reflecting} Scanning Tunneling Microscope Break Junctions (STM-BJs) as plasmonic cavities push this setup, with its advantages and disadvantages, to the extreme. In these systems, a single molecule bridges two gold electrodes separated by a nanometer-scale gap, creating both a conductive molecular junction with large loss channels and an extremely confined plasmonic nanocavity.\cite{xu2003measurement,paoletta2024plasmon} Hence, STM-BJs have served as versatile platforms for studying and manipulating single molecules, combining nanoscale field confinement with single-molecule addressability. They have enabled chemically specific conductance measurements,\cite{chen2007measurement,gorenskaia2024measurement,venkataraman2006conductance} single-molecule control,\cite{dief2023control,stone2021control} and investigations of quantum transport\cite{kim2014transport} and spin-dependent phenomena.\cite{aradhya2013tranport}

Recent work demonstrated strong light–matter coupling in such a junction, with a Rabi splitting of approximately $300 ~\mathrm{meV}$ for a single bipyridine (BP) molecule,\cite{paoletta2024plasmon} providing a new route for controllable strong light-matter engineering. This, however, raises a central question. How can such strong coupling occur at the single-molecule level in such an extremely lossy system? While one might attribute this entirely to the extreme confinement of STM-BJs, with mode volumes on the order of $1~\mathrm{to}~10~\mathrm{nm}^3$, the molecule is also directly bonded to the metallic electrodes that define the nanocavity. Beyond the general plasmonic cavity loss, this contact exposes the molecular excitation to a near continuum of metal states and potentially rapid exciton decay.

In this work, we address this question. We show that the key distinction of STM-BJs as plasmonic cavities is not only their extremely small mode volume, but the fact that they are voltage-driven rather than laser-driven. In STM-BJs, the applied bias does not merely inject carriers to form the interfacial excitation through resonant transport. It also tunes the excitation itself. Using model systems, Linear Response (LR) and Real Time (RT) Time-Dependent Density Functional Theory (TDDFT), we connect voltage-tunable electrode-molecule orbital overlap to exciton lifetime and dephasing. Across different STM-BJ-like setups, methods, and loss channels, we find a consistent optimal voltage near the resonant-transport condition. At this voltage, alignment between the electrode Fermi level and the BP-electrode LUMO resonance enables formation of the interfacial exciton while reducing its coupling to metallic loss channels. On the electromagnetic confinement side, we modify frequency-domain Maxwell simulations by including voltage-dependent gold conductivity, allowing us to probe how applied bias changes the junction mode structure and field confinement. We find that the bias preserves nanoscale mode confinement while moderately changing the availability of plasmonic modes near the relevant coupling frequency. Together, these results establish voltage-driven STM-BJs as a design principle for single-molecule strong light-matter engineering, where electrical control can be used to tune decoherence, preserve mode confinement, and access molecular-scale polaritonic functionality.

\subsection{STM-BJ as plasmonic cavities}

STM-BJ cavities differ from conventional plasmonic nanocavities in two important ways. First, the molecule is chemically bound to the metallic electrodes that define the cavity, so the relevant material excitation is interfacial rather than purely molecular. Second, the junction is voltage-driven, such that excitation occurs through so-called resonant transport rather than direct optical excitation. 

\begin{figure*}
\includegraphics[width=1.0\textwidth]{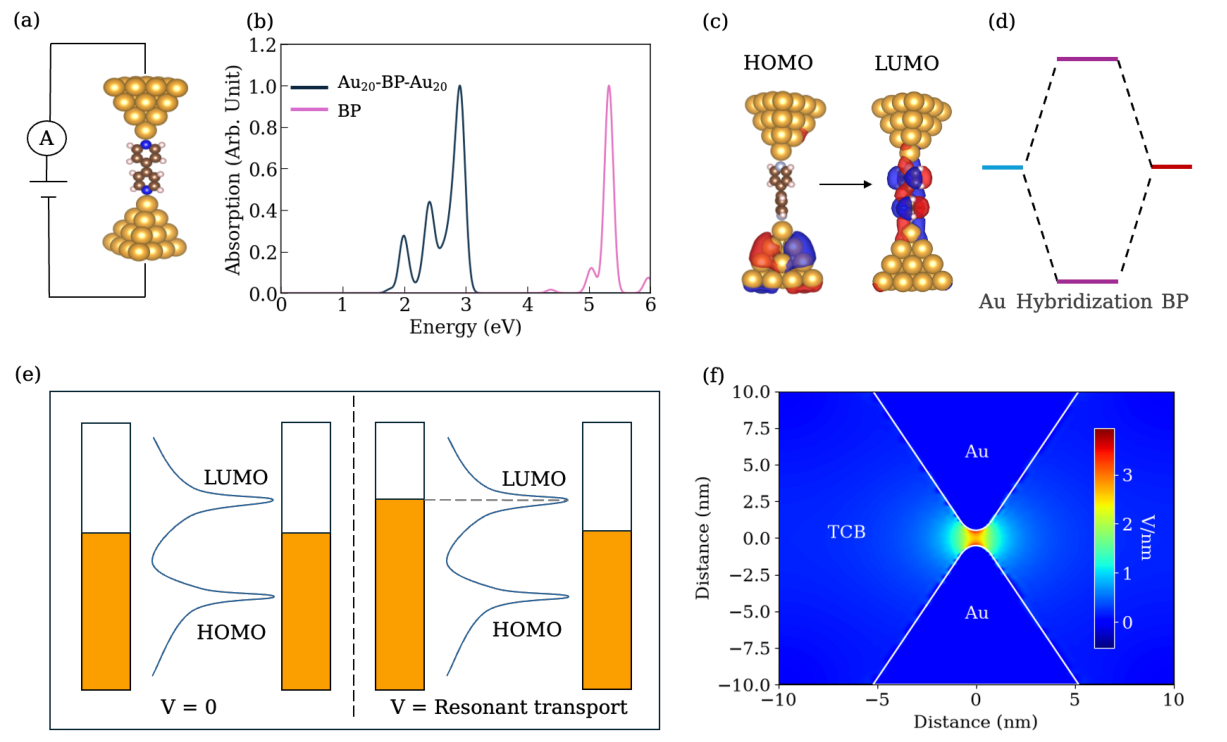}
\caption{\label{fig:overview}{(a) Schematic representation of a typical STM-BJ experimental setup.
(b) Calculated absorption spectra of BP and the Au$_{20}$--BP--Au$_{20}$ molecular junction.
(c) Molecular orbital character analysis of the HOMO--LUMO transition in the Au$_{20}$--BP--Au$_{20}$ system.
(d) Schematic illustration of orbital hybridization between BP and the Au$_{20}$ electrodes.
(e) Energy-level alignment diagram of the STM-BJ, showing the relative positions of the molecular orbitals with respect to the Fermi level of the electrodes.
(f) Electric-field distribution within the STM-BJ nanocavity under an applied bias of $2~\mathrm{V}$. 
}}
\label{fig:stm_bj_specta}
\end{figure*}

The first distinguishing feature is the interfacial character of the excitation. Fig.~\ref{fig:overview}(a) shows a schematic of a single BP molecule bridging two voltage-biased gold electrodes. Following Ref.,~\cite{paoletta2024plasmon} Fig.~\ref{fig:overview}(b) compares the LR-TDDFT calculated absorption spectrum (see Computational Details) of isolated BP (pink) with that of the Au$_{20}$-BP-Au$_{20}$ junction (black). The isolated BP has its lowest bright excitation near $5~\mathrm{eV}$ and therefore cannot couple to the gold plasmon mode near $2~\mathrm{eV}$. However, once the gold electrodes are accounted for explicitly, the junction does support low-energy transitions near the plasmon resonance. As shown previously in Ref.,~\cite{paoletta2024plasmon} the corresponding orbital structure confirms that these transitions are interfacial in character, arising from hybridization between BP molecular orbitals and Au electrode states, with electron density localized primarily on BP and the corresponding hole localized on the Au electrode (Fig.~\ref{fig:overview}(c--d)).

The second distinguishing feature is the mechanism by which this excitation is generated under applied bias. In optically driven cavities, a laser resonant with the material transition promotes an electron from an occupied molecular state into an unoccupied molecular state. In voltage-driven STM-BJs, by contrast, excitation occurs through resonant transport. Under an applied bias, the electrode Fermi levels shift relative to the molecular levels (Fig.~\ref{fig:overview}(e)). When one of them aligns with the BP-electrode LUMO, electrons can be injected from the metal into the molecule. Together, the electron in the BP-electrode LUMO and the hole in the Au electrode form the interfacial exciton identified above. At the same time, the two gold electrodes define a plasmonic nanocavity. The nanometer-scale gap supports enhanced fields in the junction region (Fig.~\ref{fig:overview}(f)).

Thus, the STM-BJ combines the three ingredients needed for strong coupling in this system: the molecule-electrode contact creates the interfacial exciton, the applied bias generates it through resonant transport, and the gold tips provide the confined plasmonic modes that couple to it.

\section*{Results}

To reiterate, the central question of this work is how an STM-BJ cavity, despite being a highly lossy metal-molecule-metal system, can exhibit a Rabi splitting of about $300~\mathrm{meV}$ at the single-molecule level.\cite{paoletta2024plasmon} This suggests that voltage-driven plasmonic cavities differ from laser-driven plasmonic cavities not only in how the excitation is generated, but also in how the excitation decays once it is formed and how the electromagnetic environment is modified under bias. We therefore analyze the problem from two complementary perspectives: the material side, which controls the interfacial excitation, its lifetime, and dephasing, and the electromagnetic confinement side, which controls the modes of the junction.

We first consider the material side, where the interfacial excitation involves both BP and Au states. Hence, its coupling to metallic loss channels is controlled by the Au-BP orbital overlap, where a possible bias-dependent change in the orbital overlap could therefore directly affect the population lifetime. To make this connection explicit, we represent the interfacial exciton with a minimal molecule-metal model. Following the Newns-Anderson and Bixon-Jortner models, \cite{Anderson1961Localized, Newns1969Self, Csaszar20220Rotational, Ghan2023Interpreting} which describe how a localized molecular state hybridizes with a continuum of metal states, we denote the BP-like state by $|a\rangle$ and an Au-like electrode state by $|k\rangle$. The total Hamiltonian is written as

\begin{equation}
H=H_0+V
\end{equation}
where $H_0$ describes the uncoupled Au and BP states, and $V$ describes the coupling between them. In the ${|k\rangle,|a\rangle}$-basis, the Hamiltonian matrix elements become

\begin{equation}
H=\left[\begin{array}{cc}
 \epsilon_k& g_{c}^* \\
  g_{c} & \epsilon_a
\end{array}\right]
\end{equation}
where $\epsilon_k$ is the energy of the Au-like state, $\epsilon_a$ is the energy of the BP-like state and $g_c$ is the molecule-metal coupling defined as
\begin{equation}
g_{c}=\langle a| V|k\rangle.
\end{equation}

Then, using Fermi's golden rule, we can describe the decay into the nearby metal-state continuum, where the transition rate $w_{\mathrm{T}}$ is given by

\begin{equation}
w_{\mathrm{T}}=2 \pi|\langle k| \hat{V}| a\rangle|^ 2 \Pi\left(E_k\right)=2 \pi {g_{c}}^2 \frac{1}{\delta}.
\end{equation}
Here $\Pi(E_k)$ is the density of  Au-states at energy $E_k$ with $\Pi(E_k)=1/\delta$, where $\delta$ is the energy spacing between neighboring Au-states. Finally, the probability of finding the system in an excited state  $|\phi\rangle$ at time $t$ can be written as
\begin{equation}
|\langle\phi \mid \Psi(t)\rangle|^2=\mathrm{e}^{-w_{\mathrm{T}} t} = \mathrm{e}^{-2 \pi {g_c}^2 \frac{1}{\delta} t}
\label{eq6}
\end{equation}
where $|\Psi(t)\rangle$ is the time-dependent coupled molecule-metal state. The full derivation is provided in Sec.~S1 of the Supporting Information.

\begin{figure*}
\includegraphics[width=1.0\textwidth]{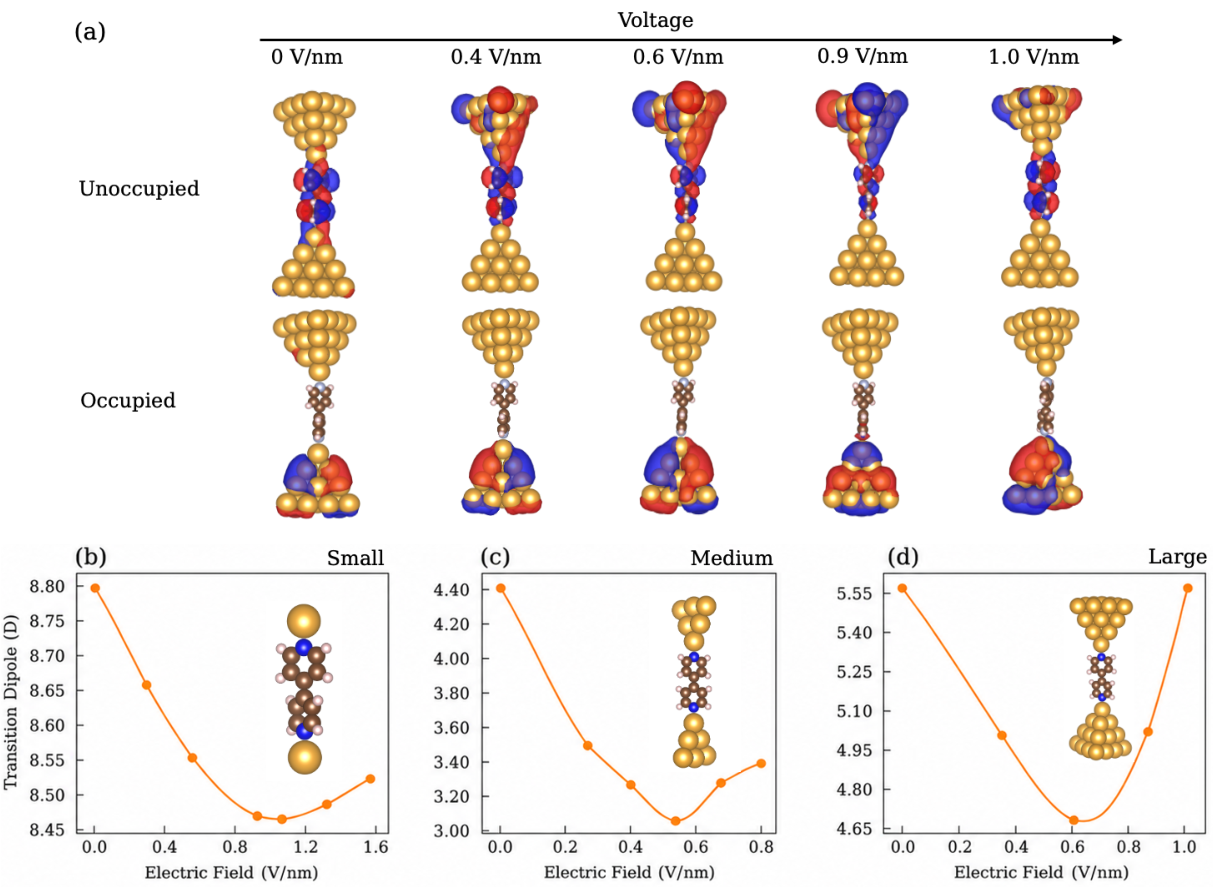}
\caption{\label{fig:trasition_dipole}(a): Molecular orbitals of the Au$_{20}$–BP–Au$_{20}$ junction under applied electric fields from $0$ to $1~\mathrm{V/nm}$, showing the bias-dependent evolution of the occupied and unoccupied orbitals. (b--d): LR-TDDFT transition dipole moment as a function of applied electric field for the small Au$_1$–BP–Au$_1$, medium Au$_6$–BP–Au$_6$, and large Au$_{20}$–BP–Au$_{20}$ junction models. All three systems exhibit a minimum at an intermediate field.}
\end{figure*}

Eq.~\ref{eq6} shows that the population lifetime is controlled by two quantities: the density of available Au states, $\Pi(E_k)=1/\delta$, and the molecule-metal coupling $g_c$. Since the gold tip structure does not change at a given set of measurements, it is reasonable to treat $\Pi(E_k)$ as approximately constant. Under this assumption, changes in the exciton lifetime are governed primarily by $g_c$. This is the essential feature of an interfacial exciton. Since $g_c$ couples the BP-like component of the excitation to Au-like electrode states, it is directly determined by the Au-BP orbital overlap. The same interfacial character that gives rise to the charge-transfer transition dipole therefore also sets the strength of the metallic loss channel. Eq.~\ref{eq6} thus connects changes in Au-BP orbital overlap to changes in the population lifetime. Fig.~\ref{fig:trasition_dipole}(a) provides a first indication of this bias dependence from the LR-TDDFT orbitals of the Au$_{20}$-BP-Au$_{20}$ junction (top panel). The Au-BP orbital overlap is similar at $0$ and $1~\mathrm{V/nm}$ but is minimized near $0.6~\mathrm{V/nm}$, suggesting that the molecule-electrode hybridization can be modulated by the applied bias.

To quantify this bias dependence systematically, we use LR-TDDFT, solving the time-dependent Kohn-Sham equations within the Casida formalism. All geometry optimizations and LR-TDDFT calculations were performed using ORCA 5.0.2 at the B3LYP/def2-TZVP level of theory\cite{neese2022}. We consider three junction models of increasing complexity: a minimal Au$_1$-BP-Au$_1$ model (Fig.~\ref{fig:trasition_dipole}(b)), an intermediate Au$_6$-BP-Au$_6$ model (Fig.~\ref{fig:trasition_dipole}(c)), and a large Au$_{20}$-BP-Au$_{20}$ model (Fig.~\ref{fig:trasition_dipole}(d)). For all cases, we find a clear parabolic-like trend with an optimal electric field where the orbital overlap and transition dipole are minimized, at $1.0~\mathrm{V/nm}$, $0.5~\mathrm{V/nm}$, and $0.6~\mathrm{V/nm}$ for the small, medium, and large models, respectively. Thus, although the absolute optimal field depends on junction size and geometry, interestingly the same qualitative behavior appears in all models.

This shows that the Au-BP orbital overlap is bias-dependent and reaches a minimum at an intermediate field. Since this overlap controls the transition dipole and the molecule-metal coupling $g_c$, the minimum marks the field where the interfacial excitation is most weakly coupled to Au-like loss channels. Thus, near the resonant-transport condition, the same bias that enables exciton formation also minimizes coupling to the Au environment. Connecting this result back to Eq.~\ref{eq6}, the reduced molecule-metal coupling at this field implies slower population decay and therefore a longer interfacial-exciton lifetime.

\begin{figure*}
\includegraphics[width=1.0\textwidth]{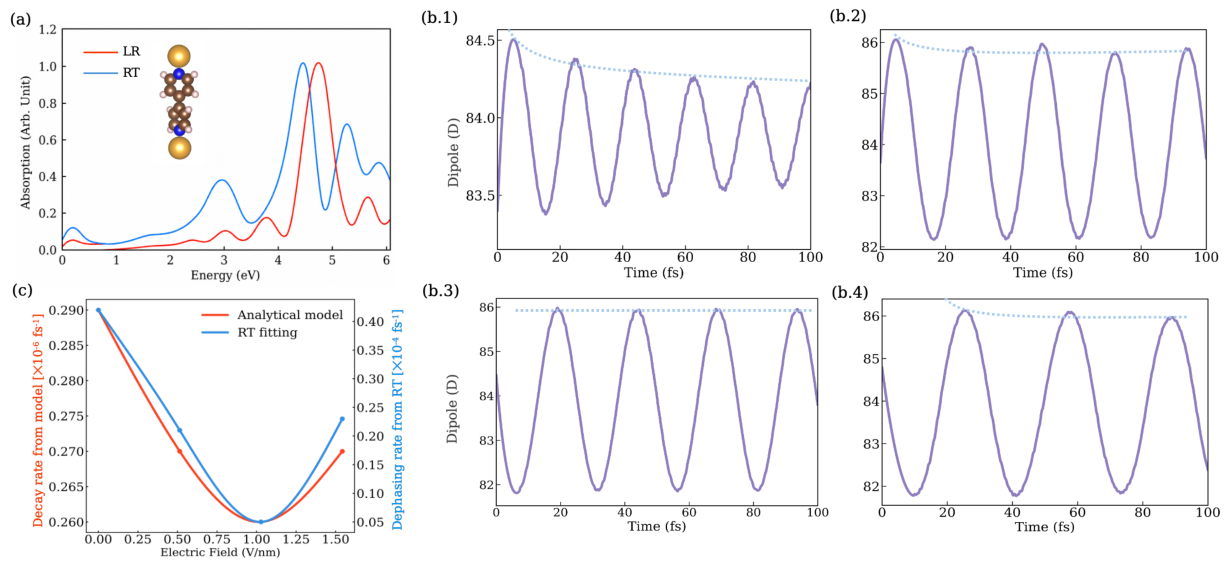}
\caption{\label{fig:wide}(a) Comparison of the absorption spectra calculated using LR-TDDFT and RT-TDDFT. 
(b.1--b.4) Time-varying RT-TDDFT dipole moments during the first 100 fs under external electric field strengths of 0, 0.6, 1.0, and 1.5 V/nm, respectively. The oscillation envelopes were fitted to determine the dephasing behavior. 
(c) Comparison between the decay rates predicted by the analytical model and the dephasing rates obtained from fitting the RT-TDDFT dipole envelopes.
} 
\end{figure*}

The LR-TDDFT results provide a static electronic-structure signature of bias-dependent lifetime tuning. We next examine whether this static trend has a dynamical counterpart. To this end, we perform RT-TDDFT simulations to follow the induced dipole dynamics after excitation, i.e., the electronic dephasing.\cite{Peng2015RealTime,Bhasin2025Plasmon} Here, we use the minimal Au$_{1}$-BP-Au$_{1}$ junction as a representative model, since the LR-TDDFT calculations already show a field-dependent trend in this system while keeping the substantially more expensive RT-TDDFT propagation computationally tractable. We perform RT-TDDFT calculations using NWChem with a B3LYP functional and def2-TZVP basis set \cite{becke1988exchange,becke1993b3lyp,lee1988lyp,weigend2005basis}. A comparison of the absorption spectra obtained from LR- and RT-TDDFT is shown in Fig.~\ref{fig:wide}(a). The interfacial excitation is then prepared by applying an electric field along the Au$_{1}$-BP-Au$_{1}$ junction axis for $40~\mathrm{fs}$, with the frequency tuned to the charge-transfer excitation energy of $3~\mathrm{eV}$. We then apply a static electric field for $15~\mathrm{fs}$ before turning off all external fields and monitoring the subsequent field-free dipole dynamics. We consider static fields of $0~\mathrm{V/nm}$, corresponding to the laser-driven reference case, $0.6~\mathrm{V/nm}$, below the resonant-transport condition, $1.0~\mathrm{V/nm}$, near the resonant-transport condition, and $1.5~\mathrm{V/nm}$, above the resonant-transport condition. The resulting dipole dynamics are shown in Fig.~\ref{fig:wide}(b.1-b.4) for the different field-prepared cases. Here, the purple lines show the field-free dipole dynamics after all external fields are turned off, and the blue dashed lines indicate the decay. For the latter, we obtain the oscillation envelope by fitting the peak values with a single-exponential decay (see also SI Sec.~S2.C).

Consistent with the LR-TDDFT results, the RT-TDDFT dynamics show a nonmonotonic dependence on the applied field. In the laser-driven reference case, corresponding to $0~\mathrm{V/nm}$, the fitted dipole-envelope decay rate is $0.42\times10^{-4}~\mathrm{fs}^{-1}$. This rate decreases to $0.21\times10^{-4}~\mathrm{fs}^{-1}$ below the resonant-transport condition and reaches a minimum of $0.05\times10^{-4}~\mathrm{fs}^{-1}$ at the resonant condition $1.0~\mathrm{V/nm}$. Increasing the field further restores faster dephasing, with a rate of $0.23\times10^{-4}~\mathrm{fs}^{-1}$ at $1.5~\mathrm{V/nm}$. Since the present RT-TDDFT simulations describe only the isolated electronic response, they do not account for additional decoherence mechanisms associated with energy dissipation, electron–phonon interactions, or coupling to the surrounding environment. Consequently, the absolute dephasing rates are not expected to quantitatively reproduce experimental coherence times. However, the simulated field dependence captures the intrinsic electronic contribution to dephasing and reveals the qualitative trend that resonant transport conditions enhance electronic coherence, whereas off-resonant fields promote faster decoherence.

This trend is summarized in Fig.~\ref{fig:wide}(c). The blue curve shows the decay rates extracted from the RT-TDDFT dipole-envelope fits, while the red curve shows the population decay rates predicted from the LR-TDDFT transition dipoles using Eq.~\ref{eq6} for the same Au$_1$-BP-Au$_1$ junction. Although the absolute magnitudes should not be compared directly, since the two quantities describe different decay channels, both curves show the same voltage dependence and reach their minimum at $1.0~\mathrm{V/nm}$. This agreement supports the interpretation that the static LR-TDDFT transition-dipole minimum and the dynamical RT-TDDFT dephasing minimum originate from the same underlying field-dependent Au-BP coupling.

Taken together, these results provide a simple physical picture of voltage-driven lifetime tuning in the STM-BJ. The applied bias first enables formation of the interfacial exciton through resonant transport, when the electrode Fermi level aligns with the BP-electrode LUMO. However, the same bias also changes the Au-BP orbital overlap that controls the coupling of this excitation back to metallic loss channels. This picture is supported by both the static LR-TDDFT transition-dipole minimum and the RT-TDDFT dipole-envelope decay minimum, which occur at the same resonant-transport condition.

Having established how the applied voltage modifies the interfacial excitation, we next consider the electromagnetic confinement part of the strong-coupling problem. Here, we examine how the applied bias modifies the confined mode volume and count in the STM-BJ. This requires a modification of the usual optical mode calculation, which treats the cavity as optically driven, such that the material response is encoded through the complex refractive index of the metal. In a voltage-driven STM-BJ, however, the applied bias can also modify the electronic response of the gold electrodes through changes in their conductivity. The Maxwell problem, therefore, contains not only the usual optical constants but also a voltage-dependent conductivity term. To account for the voltage-dependent electromagnetic response, we solve the time-harmonic Maxwell equations using COMSOL Multiphysics version 6.2 \cite{comsol62} with boundary conditions set by the STM-BJ geometry (see also Supporting Information Sec.~S4). We obtain the confined electric field by solving the time-harmonic Maxwell equation.\cite{PrinciplesofNano-Optics}

\begin{equation}
\nabla \times \mu^{-1}_r (\nabla \times \mathbf{E(r)}) - k_0^2 \left(\epsilon_r - \frac{i \sigma(V_{bias})}{\omega \epsilon_0}\right)\mathbf{E(r)} = 0
\label{EqMax}
\end{equation}
where $\mu_r$ is the relative permeability, $\epsilon_r$ is the relative permittivity, $\sigma$ is the conductivity, $\omega$ is the frequency, $k_0 = \omega \sqrt{\mu_0 \epsilon_0}$, $\mathbf{E(r)}$ is the electric field, and $V_{bias}$ is the applied bias. The relative permittivity is obtained from the complex refractive index through $\epsilon_r = (n - ik)^2$. The voltage dependence is included through the Drude-model expression for the conductivity, \cite{SolidStatePhysics, Datta_2005}

\begin{equation}
\sigma(V_{bias}) = -\frac{n_eq^2d}{m} \left( \frac{1}{\sqrt{\frac{2E_f \pm qV_{bias}}{m}}}\right)
\label{volt_cond}
\end{equation}
where $n_e$ is the free-electron density, $q$ is the elementary charge, $d$ is the mean free path, $m$ is the electron mass, and $E_f$ is the Fermi energy of gold. The full derivation is provided in Sec.~S3 of the Supporting Information.

\begin{figure}
\includegraphics[width=0.5\textwidth]{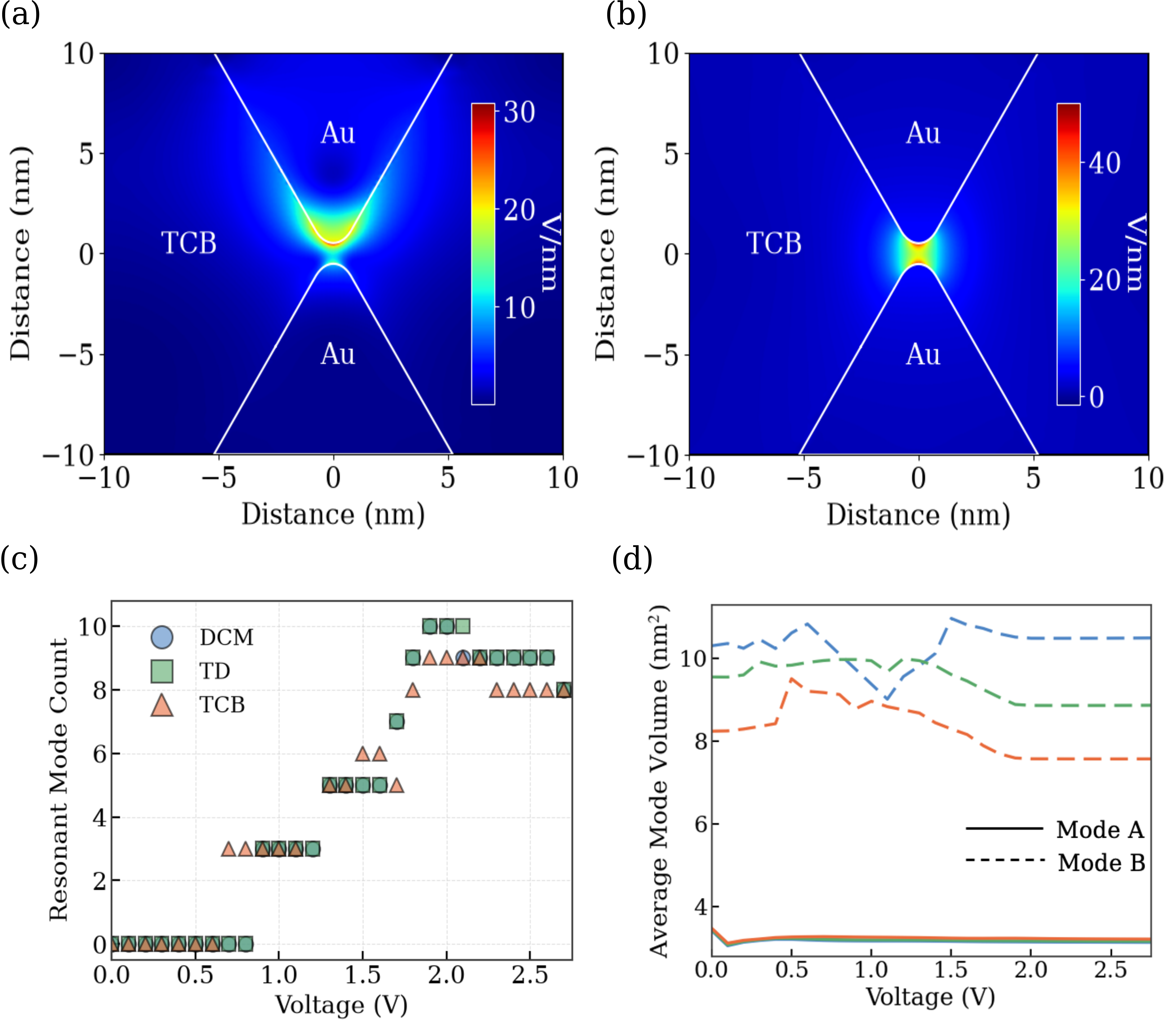}
\caption{\label{fig:photon} (a--b): Electric field distributions of two exemplary resonant plasmon modes A and B in a voltage-driven STM-BJ. (c): Resonant Mode Count calculated in DCM, TD, and TCB solvents with bias ranging from 0~V to 2.75~V. (d): Mode volumes of the exemplary resonant plasmon modes A and B as a function of bias.}
\end{figure}

We investigate the mode volume and count under applied bias in three experimentally motivated solvent environments, dichloromethane (DCM), 1,2,4-trichlorobenzene (TCB), and tetradecane (TD).

We first consider the mode volume under bias, which we define as
\begin{equation}
V_m = \frac{\int \epsilon(r) |\mathbf{E(r)}|^2 dV}{\max\left(\epsilon(r) |\mathbf{E(r)}|^2\right)},
\end{equation}
where $\epsilon(r)$ is the position-dependent permittivity. It is important to note that we divide the STM-BJ into three regions: the top tip, junction gap, and bottom tip, and evaluate the numerator and denominator region by region. For each region $i$, with $i\in\{\mathrm{top},\mathrm{gap},\mathrm{bottom}\}$, we define
$I_i=\int_i \epsilon_i(r)|\mathbf{E(r)}|^2 dV$ as the integrated electric-field density in that region and
$M_i=\max_{r\in i}\left[\epsilon_i(r)|\mathbf{E(r)}|^2\right]$ as the maximum electric-field energy density in the same region. The numerator of the mode-volume expression is then $I_{\mathrm{top}}+I_{\mathrm{cav}}+I_{\mathrm{bottom}}$, while the denominator is obtained from $M_{\mathrm{tot}}=\max(M_{\mathrm{top}},M_{\mathrm{cav}},M_{\mathrm{bottom}})$. Fig.~\ref{fig:photon}(d) shows the average mode volumes of two representative resonant modes for the three solvent environments, and Fig.~\ref{fig:photon}(a--b) show the electric field distributions of the two representative resonant modes. The absolute mode volume differs between the two modes, reflecting their different field profiles. Interestingly, however, for a given mode, the mode volume remains largely unchanged as the applied bias is varied. Thus, even when voltage-dependent conductivity is included, the extreme spatial confinement of the resonant modes is preserved.

Next, we turn to the Resonant Mode Count (RMC) as a measure of voltage-dependent mode availability. For each applied bias, we evaluate the number of resonant modes whose frequencies fall within a chosen spectral window (see also Supporting Information Sec.~S4), providing a coarse indicator of the cavity modes that dominantly contribute to the light-matter coupling. Fig.~\ref{fig:photon}(c) shows the voltage-dependent RMC for the three solvent environments. For all solvents, the mode count remains small at low bias and begins to increase only after a solvent-dependent onset voltage. With increasing bias, the RMC increases, reaches a maximum around $2~\mathrm{V}$, and then decreases slightly or saturates at higher voltage. This trend is the same for all three solvents, suggesting that applied bias produces a moderate change in the availability of plasmonic modes near the gold plasmon frequency.

Thus, the electromagnetic confinement is only weakly affected by applied bias, with the mode volume remaining nearly unchanged and the mode availability changing only to a limited extent.

\subsection*{Conclusion and Outlook}

In this work, we identify voltage as the key control parameter that allows STM-BJ plasmonic cavities to sustain strong light-matter coupling at the single-molecule level. To this end, we show how applied bias influences exciton formation, coupling to loss channels, mode volume, and plasmonic mode availability. We find that in STM-BJs, voltage does not only form the interfacial excitation through resonant transport, but also optimizes the excitation by minimizing its coupling to metallic loss channels. At the same time, the applied bias preserves the extreme mode confinement of the junction while only moderately changing the availability of plasmonic modes. This combination of voltage-controlled decoherence tuning and extreme mode confinement establishes STM-BJ as a new design principle for strong light-matter engineering.

More broadly, this sets the stage for using STM-BJ strong coupling as a new route to probe and control molecular-scale phenomena. STM-BJs are already powerful platforms for single-molecule control, chemically specific conductance measurements, and quantum transport. Strong light-matter coupling now adds a new handle to this toolbox. Unlike most cavity platforms, STM-BJs do not only provide access to optical spectra through electroluminescence, but they also provide direct access to single-molecule conductance. This makes it possible to investigate how strong coupling modifies charge transport, conductance fingerprints, and chemical dynamics in a single-molecule junction. At the same time, the enhanced electronic coherence identified here, together with the mechanically confined molecule, open interesting directions for molecular optoelectronics and quantum technologies enabled by strong light-matter coupling. Hence, voltage-driven STM-BJ plasmonic cavities set the stage for studying strong light-matter coupling not only through spectra, but through transport, chemistry, and quantum coherent molecular functionality.

\section*{Computational Details}

The tip--molecule--tip geometry was first optimized to establish the junction structure used in both simulation frameworks, i.e., the electronic excitations of the molecular junction (matter) and the electromagnetic confinement of the plasmonic cavity (light). Here, two gold pyramids were bridged by a 4,4'-bipyridine molecule. The geometry was optimized using ORCA 5.0.2\cite{neese2022} at the B3LYP/def2-TZVP level of theory.\cite{becke1988exchange,becke1993b3lyp,lee1988lyp,weigend2005basis} Each gold pyramid consisted of 20 gold atoms and was optimized layer by layer prior to adding the BP molecule for the final geometry optimization.

\noindent\paragraph{Matter}

The LR-TDDFT calculations were performed within the Casida formalism\cite{casida1995tddft} using the ORCA 5.0.2 package\cite{neese2022}, including the 200 lowest excited states. The resulting excitation energies and transition dipole moments were used as input parameters for the analytical model (Eq.~\ref{eq6}) to evaluate the population decay rates into the metallic continuum. Additional computational details are provided in Sec.~S1 of the Supporting Information. The RT-TDDFT calculations were performed using NWChem\cite{lopata_modeling_2011} at the B3LYP/def2-TZVP level of theory.\cite{becke1988exchange,becke1993b3lyp,lee1988lyp,weigend2005basis} The time-dependent Kohn--Sham orbitals were propagated following resonant excitation (40~fs) and subsequent application of a static electric field (15~fs). The induced dipole moment was smoothed using a third-order Savitzky--Golay filter,\cite{Savitzky1964Smoothing} and the oscillation maxima were fitted to a single-exponential decay to extract the dephasing rates. Further details are provided in Sec.~S2 of the Supporting Information.

\noindent\paragraph{Light}

The same optimized tip--molecule--tip geometry was used to define the junction geometry for the electromagnetic confinement calculations, yielding a tip-gap size of 1.032 nm. All electromagnetic simulations were performed using the Wave Optics Module in COMSOL Multiphysics version 6.2.\cite{comsol62} The frequency-dependent relative permittivity of gold, $\epsilon_r(\omega)$ (Eq.~\ref{EqMax}) was determined from the complex refractive index obtained in Ref.~\cite{Babar:15}. The free-electron density and Fermi energy of gold were set to $n_e=5.90 \times 10^{28}~\mathrm{m}^{-3}$ and $E_f=5.53~\mathrm{eV}$, respectively.\cite{SolidStatePhysics} The mean free path $d$ of gold electrons was treated as an effective voltage-dependent parameter in the gold tip, decreasing from an unbiased $d=2.5$~nm to $3$~\AA{} as the voltage approaches 2~V.\cite{legoas2002origin} This voltage dependence was implemented using COMSOL's piecewise cubic interpolation scheme.\cite{karna2023direct,leyland2008energy} The surrounding dielectric environment was described using wavelength-dependent refractive indices for DCM\cite{bertie1995infrared} and TCB,\cite{myers2018accurate} while a constant refractive index of $n_{\mathrm{TD}}=1.4290$ was used for TD.\cite{haynes2016crc} The average mode volume was obtained by averaging the mode volumes of resonant modes with the same electric-field distribution geometry across the applied biases. Further details of the electromagnetic simulation workflow are provided in Sec.~S4 of the Supporting Information.

\section*{Acknowledgments}

This work was supported by the Simons Center for Computational Physical Chemistry (SCCPC) at NYU (SF Grant No. 839534). N.M.H. acknowledges support from the U.S. DOE, Office of Science, BES (Early Career Award No. DE-SC0026328), as managed by the CPIMS program. Additional support was provided in part through NYU IT High Performance Computing resources, services, and staff expertise. Simulations were partially executed on resources supported by the SCCPC. The authors thank Irén Simkó for the helpful discussion.

\section*{Supporting Information}
Additional computational details, theoretical derivations, and simulation protocols.

\section*{Data Availability Statement}
Data are available from the authors upon reasonable request.

\section*{Conflict of Interest Disclosure}
The authors declare no competing financial interest.

\bibliography{apssamp}

\end{document}


\title{
Decoherence Tuning in Voltage-Driven Plasmonic Cavities}

\author{Yuchen Wang\textsuperscript{$\dagger$}}
\altaffiliation{Current affiliation: State Key Laboratory of Precision and Intelligent Chemistry, University of Science and Technology of China, Hefei, China}
\affiliation{Department of Chemistry, New York University, New York, NY 10003}
\affiliation{The Simons Center for Computational Physical Chemistry, New York University, New York, NY 10003}

\author{Oliver Tan}
\altaffiliation{Y.W. and O.T. contributed equally to this work.}
\affiliation{Department of Chemistry, New York University, New York, NY 10003}

\author{Norah M. Hoffmann}
\altaffiliation{Corresponding author: \href{mailto:norah.hoffmann@nyu.edu}{norah.hoffmann@nyu.edu}}
\affiliation{Department of Chemistry, New York University, New York, NY 10003}
\affiliation{The Simons Center for Computational Physical Chemistry, New York University, New York, NY 10003}
\affiliation{Department of Physics, New York University, New York, NY 10003}

\maketitle

\tableofcontents
\setcounter{tocdepth}{2} 

\section{Derivation of the Two-Level Analytical Model}

Considering a two-component system, we define a discrete eigenstate representing the molecular state $|\phi\rangle$ with eigenvalue $E_\phi$:
\begin{equation}
\hat{H}_0|\phi\rangle = E_\phi|\phi\rangle
\end{equation}
and a discretized continuum of states $|k\rangle$ with eigenvalues
\begin{equation}
E_k = k\delta,
\end{equation}
where $k \in \mathbb{Z}$ labels the discretized continuum and $\delta$ is the energy spacing between adjacent states.
We assume a constant, energy-independent coupling between the discrete and continuum states,
\begin{equation}
\langle \phi | \hat{V} | k \rangle = v.
\end{equation}
Applying Fermi’s golden rule, the transition (decay) rate is given by
\begin{equation}
w_{\mathrm{T}} = 2\pi |\langle k | \hat{V} | \phi \rangle|^2 \rho(E_\phi),
\end{equation}
where $\rho(E_\phi)$ is the density of states, with $\rho(E_\phi)=1/\delta$ in the wide-band limit of a uniformly discretized continuum, yielding
\begin{equation}
w_{\mathrm{T}} = \frac{2\pi v^2}{\delta}.
\end{equation}
The full system Hamiltonian is
\begin{equation}
\hat{H}|\psi_\mu\rangle = (\hat{H}_0 + \hat{V})|\psi_\mu\rangle = E_\mu|\psi_\mu\rangle,
\end{equation}

and the projection of the dressed eigenstate onto the discrete state can be obtained from the normalization of the Newns–Anderson eigenstates, \cite{Anderson1961Localized, Newns1969Self}
\begin{equation}
|\langle \phi | \psi_\mu \rangle|^2 =
\frac{1}{2\pi}
\frac{w_{\mathrm{T}}}{(E_\mu - E_\phi)^2 + (w_{\mathrm{T}}/2)^2}.
\end{equation}
Expanding the initial state $|\Psi(0)\rangle = |\phi\rangle$ in the eigenbasis of $\hat{H}$ gives
\begin{equation}
|\Psi(0)\rangle =
\sum_\mu \langle \psi_\mu | \phi \rangle |\psi_\mu\rangle,
\end{equation}
and the time-evolved state
\begin{equation}
|\Psi(t)\rangle =
\sum_\mu \langle \psi_\mu | \phi \rangle e^{-iE_\mu t} |\psi_\mu\rangle.
\end{equation}
The survival amplitude is \cite{Csaszar20220Rotational}
\begin{equation}
\langle \phi | \Psi(t) \rangle
\approx
\int_{-\infty}^{\infty}
\frac{w_{\mathrm{T}}}{2\pi}
\frac{e^{-iEt}}{(E - E_\phi)^2 + (w_{\mathrm{T}}/2)^2}
\, dE,
\end{equation}
where we have used the continuum limit with a flat density of states. Evaluating the integral yields
\begin{equation}
\langle \phi | \Psi(t) \rangle =
e^{-iE_\phi t - w_{\mathrm{T}} t/2}, \quad t \ge 0.
\end{equation}
Therefore, the survival probability is
\begin{equation}
|\langle \phi | \Psi(t) \rangle|^2 =
\exp\left(-w_{\mathrm{T}} t\right)
=
\exp\left(-\frac{2\pi v^2}{\delta} t\right).
\end{equation}

The molecule--metal coupling strength was determined from the transition dipole moment obtained using LR-TDDFT. It should be noted that the LR-TDDFT transition dipole was used only to parameterize the relative field dependence of the molecule--metal coupling, according to $v(F)/v(F_{\rm ref})=|\mu(F)|/|\mu(F_{\rm ref})|$, where $F$ denotes the applied electric-field strength and $F_{\rm ref}=0$ is the zero-field reference condition. Assuming a field-independent continuum spacing $\delta$, the rates were calculated as $w_T(F)=w_{T,\rm ref}[|\mu(F)|/|\mu(F_{\rm ref})|]^2$. Therefore, the reported values are calibrated rates in fs$^{-1}$, while the LR-TDDFT transition dipoles determine only their relative field dependence.
Taking the Au$_{1}$--BP--Au$_{1}$ junction as an example and assuming an electronic energy spacing of 3~eV, the analytical model predicts decay rate constant of $2.88 \times 10^{-7}\,\mathrm{fs^{-1}}$, $2.70 \times 10^{-7}\,\mathrm{fs^{-1}}$, $2.65 \times 10^{-7}\,\mathrm{fs^{-1}}$, and $2.68 \times 10^{-7}\,\mathrm{fs^{-1}}$ under applied static electric fields of $0~\mathrm{V\,nm^{-1}}$, $0.6~\mathrm{V\,nm^{-1}}$, $1.0~\mathrm{V\,nm^{-1}}$, and $1.5~\mathrm{V\,nm^{-1}}$, respectively. Because the excited-state lifetime is given by $\tau = 1/k$, a smaller decay rate constant corresponds to a longer excited-state lifetime.

\section{Details of Real-Time Time-Dependent Density Functional Theory Simulations}

\subsection{Benchmarking}

Unlike LR-TDDFT, which directly computes discrete excitation energies, RT-TDDFT obtains the full frequency-dependent electronic response by Fourier transforming the time-dependent dipole moment following an external perturbation. In the weak-field limit, the absorption spectrum obtained from RT-TDDFT using a delta-kick perturbation is expected to be equivalent to that obtained from LR-TDDFT.

To benchmark the consistency between the two implementations, RT-TDDFT calculations were performed on the Au-BP-Au junction using a delta-kick perturbation, and the resulting absorption spectrum was compared with the corresponding LR-TDDFT spectrum calculated for the same system.\onlinecite{WEERAWARDENE201827} The comparison is presented in Fig.~3 in the main text.

As shown in Fig.~3, both approaches predict the dominant absorption peak at approximately 3.0 eV, demonstrating good agreement in excitation energies. Small differences in the absorption intensities are observed because ORCA employs the Tamm--Dancoff approximation (TDA) by default.\onlinecite{HIRATA1999291} Within the TDA, the coupling matrix in the Casida equations is neglected, such that the oscillator-strength sum rule is not strictly satisfied. Consequently, slight deviations in oscillator strengths are expected. However, the good agreement in peak positions demonstrates the consistency between the two TDDFT implementations in describing the excitation energies and supports the use of the RT-TDDFT methodology throughout this work.

\subsection{Simulation Protocol for Excitation and Relaxation Dynamics}

Two RT-TDDFT simulation protocols were employed to investigate the excitation and relaxation dynamics of the molecular junction. First, a continuous-wave (CW) electric field resonant with the charge-transfer excitation of the junction was applied for 40 fs to excite the system. The external field was then switched off, and the induced electron dynamics were monitored during the subsequent voltage-free relaxation. The same resonant CW excitation was applied for 40 fs, followed by a 15 fs static electric field to probe the effect of bias on the excited-state dynamics. All external fields were then removed, and the system was propagated under field-free conditions.

\subsection{Analysis of Dipole Relaxation Dynamics}

The decay of the induced dipole moment was quantified from the RT-TDDFT simulations. To reduce high-frequency numerical fluctuations while preserving the overall oscillation envelope, the raw dipole signal was smoothed using a Savitzky--Golay filter\cite{Savitzky1964Smoothing} with a window length of 8000 and a third-order polynomial.

The smoothed signal was subsequently interpolated onto a uniformly spaced grid to improve the accuracy of peak detection. The local maxima corresponding to each oscillation cycle were then identified and fitted to a single-exponential decay function,

\begin{equation}
D(t)=Ae^{-kt},
\end{equation}

where $A$ is the initial dipole amplitude and $k$ is the decay rate.

The fitted decay rate constants under different applied electric fields are summarized in Table~\ref{tab:tau_values_rt}.

\begin{table}[h!]
\centering
\caption{Exponential fitting parameters for the RT-TDDFT dipole relaxation dynamics.}
\label{tab:tau_values_rt}
\begin{tabular}{cccc}
\hline
Electric field (V/nm) & $A$ (Debye) & $k$ ($10^{-4}$ fs$^{-1}$) & Error ($10^{-4}$ fs$^{-1}$) \\
\hline
0.0 & 84.50 & 0.42 & 0.03 \\ \hline
0.6& 85.97 & 0.21 & 0.07 \\ \hline
1.0 & 86.06 & 0.05 & 0.02 \\ \hline
1.5 & 86.13 & 0.23 & 0.08 \\ 
\hline
\end{tabular}
\end{table}

\section{Voltage-Dependent Conductivity}

For the voltage-driven electromagnetic simulations, the applied bias shifts the Fermi energies of the two tips by $\pm \frac{1}{2}qV_{\mathrm{bias}}$.\cite{Datta_2005} We treat the bias-dependent conductivity as a correction to the zero-bias optical response of gold, which is already included through its complex refractive index. This correction is taken to be negative to account for increased electron scattering under applied bias.\cite{leyland2008energy} Starting from the DC Drude conductivity, the bias-dependent conductivity is then given by\cite{SolidStatePhysics}
\begin{equation}
    \sigma = \frac{n_eq^2\tau_e}{m}.
\end{equation}
Here, $\tau_e$ is the mean time between electron collisions and can be expressed in terms of the mean free path $d$ and electron velocity $v$, which yields
\begin{equation}
    \sigma = \frac{n_eq^2d}{mv}.
\end{equation}
The electron velocity is related to the Fermi energy through
\begin{equation}
    v = \sqrt{\frac{2E_f}{m}}.
\end{equation}
Substituting $E_f \rightarrow E_f \pm \frac{1}{2}qV_{\mathrm{bias}}$ and including the negative conductivity correction yields the bias-dependent conductivity
\begin{equation}\label{eq:volt_conduct}
    \sigma(V_{\mathrm{bias}}) = -\frac{n_eq^2d}{m}
    \left(
    \frac{1}{\sqrt{\frac{2\left(E_f \pm \frac{1}{2}qV_{\mathrm{bias}}\right)}{m}}}
    \right).
\end{equation}

\section{Resonant Mode Workflow}

In this section, we describe the step-by-step workflow for the resonant mode calculations in COMSOL Multiphysics version 6.2.\cite{comsol62}:

\begin{itemize}
    \item \textbf{Geometry:} The STM-BJ tip--tip geometry was constructed within a $10~\mathrm{nm}\times10~\mathrm{nm}$ simulation domain. Two diagonal lines extending from approximately $\pm$5 nm to $\pm$1 nm were used to construct the base of each tip. A quadratic Bezier curve was then applied to generate the tip geometry while maintaining a 1.032 nm gap, as determined from the geometry optimization. The material properties were then assigned to each region of the geometry, with gold for the tips and TCB, DCM, or TD solvents for the junction gap.
    \item \textbf{Parameters:} The refractive index, relative permeability, and relative permittivity (calculated from the refractive index) were assigned for each material. The free-electron density of gold was defined as $n_e=5.90 \times 10^{28}~\mathrm{m}^{-3}$ and the Fermi energy as $E_f=5.53~\mathrm{eV}$. \cite{SolidStatePhysics} The mean free path $d$ of gold electrons was treated as an effective voltage-dependent parameter in the gold tip, decreasing from an unbiased $d=2.5$~nm to $3$~\AA{} as the voltage approaches 2~V. This voltage-dependence was implemented using the piecewise cubic interpolation function in COMSOL Multiphysics version 6.2. \cite{comsol62} The gold refractive index was taken from Babar and Weaver, with $\mu_r=0.999998$.\cite{Babar:15,Edwards2016-fx} The wavelength-dependent refractive indices of the TCB, DCM, and TD solvents were taken from Refs.~\cite{myers2018accurate,bertie1995infrared,haynes2016crc}, respectively, while $\mu_r=1$ was assumed for all three solvents.
    \item \textbf{Modes:}  We solve the time-harmonic Maxwell equation, Eq.~(6) of the main text, with $\sigma(V_{\mathrm{bias}})$ given by Eq.~(\ref{eq:volt_conduct}), to obtain the resonant electric-field modes. The simulations were initialized with an electric field of $0~\mathrm{V/m}$, with scattering boundary conditions applied at the edges of the simulation domain. The finest available mesh was used throughout. Eigenfrequency calculations were performed using a parametric bias sweep from 0 to 10~V over frequencies near the gold plasmon frequency, $\omega_p=2.34$~eV.
    \item \textbf{Resonant Mode Count:} The electric-field distributions, resonant mode count (RMC), and average mode volumes were analyzed. The RMC was determined by counting the resonant modes within an energy window of $\pm0.17$~eV around the gold plasmon frequency at each applied bias. The average mode volume was obtained by averaging the mode volumes of resonant modes with the same electric-field distribution geometry across the applied biases.
\end{itemize}

\bibliography{apssamp}